\documentclass[10pt, aps, prc, superscriptaddress, nofootinbib, 
               amsmath, amssymb, twocolumn,
               preprintnumbers, showpacs,
               raggedbottom]{revtex4-1}

\newcommand{\figdir}{figures}     
\usepackage[arXiv]{my_paper}     
\usepackage{my_acronyms}          

\usepackage[utf8]{inputenc}

\DeclareMathOperator{\sinc}{sinc}
\begin{document}
\title{Use of the Discrete Variable Representation Basis in Nuclear Physics}

\author{Aurel Bulgac}
\affiliation{Institute for Nuclear Theory, University of Washington,
  Seattle, Washington 98195--1550 \mysc{usa}}

\author{Michael McNeil Forbes}
\affiliation{Department of Physics, University of Washington,
  Seattle, Washington 98195--1560 \mysc{usa}}
\affiliation{Institute for Nuclear Theory, University of Washington,
  Seattle, Washington 98195--1550 \mysc{usa}}
 
\date{\today}
\preprint{\upsc{NT@UW}-13-08, \upsc{INT-PUB}-13-004}

\glsresetall
\begin{abstract}
  \noindent
  The \gls{DVR} basis is nearly optimal for numerically representing wave
  functions in nuclear physics: Suitable problems enjoy exponential convergence,
  yet the Hamiltonian remains sparse. We show that one can often use smaller
  basis sets than with the traditional harmonic oscillator basis, and still
  benefit from the simple analytic properties of the \gls{DVR} bases which
  requires no overlap integrals, simply permit using various Jacobi coordinates,
  and admit straightforward analyses of the \acrlong{UV} and \acrlong{IR}
  convergence properties.
\end{abstract}


\maketitle

\glsresetall
\glsunset{FFTW}
\glsunset{MATLAB}
\glsunset{DoE}
\glsunset{RAM}

\lettrine{P}{roblems} in nuclear physics typically require solving the one-body
Schrödinger equation in three-dimensions.  Numerically representing
wavefunctions requires limiting both \gls{UV} and \gls{IR} scales: a finite
spatial resolution (i.e., a lattice) characterizes the highest representable
momenta $\Lambda$, while a finite size (i.e. a cubic box of volume $L^3$)
determines the largest physical extent.  Nuclear structure calculations are
historically dominated by the use of the \gls{HO} basis of \gls{HO} wave
functions.  The appeal of the \gls{HO} basis stems from the shape of the
self-consistent field obtained for small nuclei, which can be approximated by a
harmonic potential at small distances from the center of the nucleus.  One can
also use the Talmi-Moshinsky transformation to separate out the center-of-mass
motion in products of single particle \gls{HO} wavefunctions.  Recent efforts
have been made to determine a minimal \gls{HO} basis set, and to understand its
convergence and accuracy~\cite{Furnstahl:2012, Coon:2012}.

Here we advocate that the \gls{DVR} -- in particular the Fourier plane-wave
basis -- enjoys most of the advantages of the \gls{HO} basis, but with a
significant improvement in terms of computational efficiency and simplicity,
thereby admitting straightforward \gls{UV} and \gls{IR} convergence analyses and
implementation.

Consider wavefunctions in a cubic box of volume $L^3$ with momenta less than
$\Lambda$. The total number of quantum states in such a representation is given
by the following intuitive formula -- the ratio of the total phase space volume
to the phase space volume of a single three-dimensional quantum state:
\begin{equation}
  \label{eq:N_qs}
  \mathcal{N}_{\text{QS}} = \left( \frac{L\; 2\Lambda}{2\pi\hbar} \right)^3.
\end{equation}
One obtains the same result~\cite{Hamming:1973} using Fourier analysis: there
are exactly $\mathcal{N}_{\text{QS}}$ linearly independent functions in a cubic
3\mysc{d} box of volume $L^3$ with periodic boundary conditions and wave-vectors
less than $k_c = \Lambda/\hbar$ in each direction.  These can be conveniently
represented in the coordinate representation with $N$ equally spaced points in
each direction and lattice constant $a = \pi/k_c = \pi\hbar/\Lambda = L/N$ for a
total of $N^3 = \mathcal{N}_{\text{QS}}$ coefficients.  The maximum wave-vector
$k_c$ is simply the Nyquist frequency~\cite{Hamming:1973}; one gains nothing by
sampling the functions on intervals (``times'') finer than $a$.

The wavefunctions can also be represented in momentum space using a discrete
\gls{FFT}~\cite{FFTW}.  The momentum representation consists of
$\mathcal{N}_{\text{QS}}$ coefficients on a 3\mysc{d} cubic lattice with spacing
$2\pi\hbar/L$ and extent $-\Lambda \leq p_{x,y,z} < \Lambda$.  Using the
\gls{FFT} to calculate spatial derivatives is not only fast with $N\log N$
scaling, but extremely accurate -- often faster and more accurate than
finite-difference formulas. We use an even number of lattice points ($N=2^n$ is
best for the \gls{FFT}) and quantize the three momenta ($p_{x,y,z}=\hbar
k_{x,y,z}$)
\begin{gather}
  \begin{aligned}
    p_k &= \frac{2\pi k\hbar}{L}, &
    x_k &= ak,
  \end{aligned} \nonumber\\
  k \in \left ( -\tfrac{N}{2},-\tfrac{N}{2}+1,\dots,\tfrac{N}{2}-1 \right ).
  \label{eq:pn}
\end{gather}
The Fourier basis uses plane waves -- e.g. $\exp(ik_nx)$ in the $x$-direction --
but these can be linearly combined to form an equivalent sinc-function basis:
\begin{equation}
  \psi_k(x) = \sinc k_c(x-x_k) = \frac{\sin k_c (x - x_k)}{k_c (x-x_k)}. 
  \label{eq:sinc}
\end{equation}
This is similar to the difference between Bloch and Wannier wave functions in
condensed matter physics.  An advantage of this basis is that it is quasi-local
$\psi_k(x_l) = \delta_{kl}$ allowing one to represent external potentials as a
diagonal matrix $V_{kl} \approx V(x_k)\delta_{kl}$~[see Eq.~\eqref{eq:V}].

The plane wave basis can thus be interpreted as a periodic \gls{DVR} basis set,
which has been discussed extensively in the literature
(see~\cite{Littlejohn:2002, LCCMP:2002, LC:2002,Baye:1995, Baye:2006} and the
references therein), and one can take advantage of Fourier techniques and the
useful \gls{DVR} properties.

In general, \gls{DVR} bases are characterized by two scales: a \gls{UV} scale
$\Lambda = \hbar k_c$ defining the largest momentum representable in the basis,
and an \gls{IR} scale $L$ defining the maximum extent of the system.  In many
cases, the basis is constructed by projecting Dirac $\delta$ functions onto the
finite-momentum subspace: For example, the sinc-function basis~\eqref{eq:sinc}
is precisely the set of projected Dirac $\delta$ functions $\psi_n(x) = P_{p\leq
  \Lambda}\delta(\vect{r}-\vect{r}_\alpha)$ onto the subspace $|\vect{p}|\leq
\Lambda$~\cite{Littlejohn:2002,LCCMP:2002,LC:2002}.  (It can be non-trivial,
however, to choose a consistent set of abscissa maintaining the quasi-locality
property.) The basis thus optimally covers the region $[-L/2, L/2) \times
[-\Lambda, \Lambda)$ for each axis in phase space, and leads to an efficient
discretization scheme with exponential convergence properties.

The \gls{DVR} basis admits a straightforward analysis of the \gls{UV} and
\gls{IR} limits, allowing one to construct effective extrapolations to the
continuum and thermodynamic limits respectively.  The \gls{UV} effects may be
analyzed by simply considering the properties of the projection
$P_{p\leq\Lambda}$ used to define the basis, and the \gls{IR} limit for the
periodic basis is well understood by techniques like those derived by Beth,
Uhlenbeck, and L\"uscher~\cite{Beth:1937, Luscher:1986, *Beane:2003da}. We would
like to emphasize an additional technique here: The \gls{IR} limit is
characterized by $2\pi\hbar/L$ -- the smallest interval in momentum space
resolvable with the basis set.  For some problems, one can efficiently
circumvent this limitation by using ``twisted'' boundary conditions
$\psi(\vect{r}+\vect{L}) = \exp(\I\theta_B)\psi(\vect{r})$ or Bloch waves as
they are known in condensed matter physics.  In particular, averaging over
$\theta_B \in[0, 2\pi)$ will completely remove any \gls{IR} limitations (without
changing the basis size) for periodic and homogeneous problems, effectively
``filling-in'' the momentum states $p_n \leq p_n + \hbar\theta_B/L <
p_{n+1}$. Extensions of these formulas to the case of a box with unequal sides
is straightforward.

To demonstrate the properties of the \gls{DVR} basis, we contrast it with the
\gls{HO} basis.  The periodic \gls{DVR} basis (plane-waves) shares the ease of
separating out the center-of-mass.  In particular, one can use Jacobi
coordinates to separate out the center-of-mass motion without evaluating
Talmi-Moshinsky coefficients, leading to simpler and more transparent
implementations.  The quasi-locality of the \gls{DVR} basis offers an additional
implementation advantage over the \gls{HO} basis: one need not compute
wavefunction overlaps to form the potential energy matrix.  In contrast with the
\gls{HO} basis, the kinetic energy matrix $\mat{K}$ is no-longer diagonal, but
it has an explicit formula~\eqref{eq:DVR_K}, and is quite sparse, unlike the
potential energy operator in the \gls{HO} basis.

Consider the \gls{HO} wavefunctions with energy $E \leq \hbar\omega(N + 3/2)$:
the maximum radius and momenta are
\begin{align}
  R &= \sqrt{2N+3}\, b, &
  \Lambda &= \sqrt{2N+3}\, \frac{\hbar}{b},
  \label{eq:LP}
\end{align}
where $b=\sqrt{\hbar/m\omega}$ is the oscillator length.  For large $N$,
$N\approx R\Lambda/2\hbar$.  Thus, to expand a wavefunction with extent $2R$
containing momenta $\abs{p} < \Lambda$ requires at least
\begin{equation}
  \mathcal{N}_{\gls{HO}} = \frac{(N+1)(N+2)(N+3)}{6}
  \approx  \frac{1}{6}  \left ( \frac{R \Lambda }{2\hbar}\right )^3
  \label{eq:ho}
\end{equation}
states.  To contrast, the \gls{DVR} basis covering the required volume of phase
space~\eqref{eq:N_qs} with $L=2R$ and $\Lambda$ is
\begin{equation}
  \mathcal{N}_{\gls{DVR}}= \left ( \frac{2R\; 2\Lambda}{2\pi\hbar} \right )^3.
\end{equation}
The ratio in the limit $N\rightarrow \infty$ is thus
\begin{equation}
  \frac{\mathcal{N}_{\gls{DVR}}}{\mathcal{N}_{\gls{HO}}} 
  = \frac{384}{\pi^3} \approx 12.4.
  \label{eq:ratio1}
\end{equation}
Since these states are localized, one can further impose Dirichlet boundary
conditions, allowing functions only of the type $\sin(k_nx)$ with $k_nL = n\pi$
(instead of $\exp(ik_nx)$), thereby keeping only half of the momenta:
\begin{equation}
  \frac{\mathcal{N}_{\gls{DVR}}}{\mathcal{N}_{\gls{HO}}}=\frac{48}{\pi^3}
  \approx 1.5.
  \label{eq:ratio2}
\end{equation}
Choosing a cubic box with Dirichlet boundary conditions, sides $L=\SI{40}{fm}$,
and maximum momentum $\Lambda = \SI{300}{MeV}/c$ gives
\begin{equation}
  \mathcal{N}_{\gls{DVR}} = \left ( \frac{ L \; \Lambda }{2\pi\hbar} \right )^3
  \approx 10^3, 
\end{equation}
a somewhat surprisingly small number of states.  For symmetric states, one could
further the reduce the basis by imposing cubic symmetry, decreasing the basis
size by another factor of 8.

Finally, one can fully utilize spherical symmetry with a related Bessel-function
\gls{DVR} basis gaining a factor of $\pi/6$, and thereby besting the \gls{HO}
basis
\begin{equation}
  \frac{\mathcal{N}_{\gls{DVR}}}{\mathcal{N}_{\gls{HO}}} 
  = \frac{8}{\pi^2} \approx 0.8 <1.  
\end{equation}
In this counting, spin and isospin degrees of freedom which occur in both bases
have been omitted.
 
The Bessel-function \gls{DVR} basis set \cite{Littlejohn:2002, LCCMP:2002,
  LC:2002, Nygaard:2004} follows from a similar procedure of projecting Dirac
$\delta$ functions for the radial Schrödinger equation.  The angular
coordinates are treated in the usual manner using spherical harmonics, but the
radial wavefunctions are based on the Bessel functions (see Refs.~\cite{LC:2002,
  Nygaard:2004} for details) which satisfy the orthogonality conditions
\begin{equation}
  \int_0^{k_c}dk  \frac{2k}{k_c^2} 
  \frac{ J_\nu(kr_{\nu \alpha})J_\beta(kr_{\nu \beta}) }
       { |J_\nu'(kr_{\nu \alpha})J_{\nu \beta}'(kr_{\nu \beta})| }
       = \delta_{\alpha\beta},
\end{equation}
where $z_{\nu \alpha}= k_cr_{\nu \alpha}$ [the zeros of the Bessel functions
$J_\nu(z_{\nu \alpha})=0$] define the radial abscissa $r_{\nu, \alpha}$. The
\gls{DVR} basis set is
\begin{align}
  F_{\nu n}(r) &= \sqrt{r} J_\nu \left (\frac{z_{\nu n}r}{R} \right ), & 
  z_{\nu n} &= k_c r_{\nu n}.
\end{align}
Differential operators have simple forms in the \gls{DVR} basis
(see Refs.~\cite{Littlejohn:2002, LCCMP:2002, LC:2002} and the
codes~\cite{matlab, plots}).  In principle, a different basis (and corresponding
abscissa) should be used for each angular momentum quantum number $\nu$; In
practice, good numerical accuracy is obtained using the $\nu=0$ basis
$j_0(z_{0n}r/R)$ and the $\nu=1$ basis $j_1(z_{1n}r/R)$ respectively for even
and odd partial waves~\cite{Nygaard:2004, matlab}.  In the S-wave case, the
abscissa are simply the zeros of the spherical Bessel function
$j_0(z)=\sin(z)/z$:
\begin{align}
  z_{0n} &= n\pi, &
  r_{0n} &= \frac{n\pi}{k_c}, &
  n &= 1,2,3,\dots, N,
\end{align}
and correspond to the 1\mysc{d} basis with Dirichlet boundary conditions
mentioned earlier.  The zeros for $j_1(z)$ lie between the zeros of
$j_0(z)$. The number of \gls{DVR} functions needed to represent with exponential
accuracy a radial wavefunction is
\begin{equation}
  \mathcal{N}_{0\, \gls{DVR}} = \frac{Rk_c}{\pi},
\end{equation}
to be compared (in the limit $N\rightarrow \infty$) with 
\begin{equation}
  \mathcal{N}_{0\, \gls{HO}} = \frac{Rk_c}{4}.
\end{equation}
In the last formula we have divided by an additional factor of 2, since $N=2n+l$
changes in steps of 2.

A major drawback of the \gls{HO} wavefunctions that is rarely mentioned is that,
for modest values of $N$ and $l \neq 0$, the radial wave functions concentrate
in two distinct regions: around the inner and outer turning points of the
effective potential $V(r) = \hbar^2l(l+1)/2mr^2 + m\omega^2r^2/2$.  By adding
components with larger values of $N$, one modifies the wavefunction at both
small and large distances, leading to slow convergence.  In contrast, the
\gls{DVR} functions are concentrated around a single lattice site.  Thus, adding
more components only affects the solution in the vicinity of the additional
lattice points leaving the states largely unaffected elsewhere.

For nuclei one can gain insight with some estimates.  Cutoffs of $\Lambda=
\SI{600}{MeV}/c$ and $R=1.5\cdots 2 A^{1/3}$\si{fm} should satisfy most of the
practical requirements, leading to
\begin{subequations}
  \begin{align}
    b &= \sqrt{\frac{\hbar R}{\Lambda}} \approx 0.7 \cdots 0.8 \,
    A^{1/6} \si{fm},\\
    \hbar \omega = \frac{\hbar^2}{mb^2} &= \frac{\hbar \Lambda}{mR} 
    \approx 60 \cdots 80 \, A^{-1/3}\si{MeV}, \label{eq:omega}
  \end{align}
\end{subequations}
compared to the value $40\, A^{-1/3}\si{MeV}$ one finds in typical
monographs~\cite{Bohr:1998}. Using only half the value of $\Lambda =
\SI{300}{MeV}/c$ naturally halves the value of $\hbar \omega$.

We end with demonstrations of the \gls{DVR} method~\cite{plots}.  We start with
the harmonic oscillator problem in 1\mysc{d}
\begin{equation}
  H\phi(x)  = \left(-\frac{\hbar^2}{2m} \frac{d^2}{dx^2} 
  + \frac{k_c^2x^2}{2R^2}\right)\phi(x) = E \phi(x),
  \label{eq:H_HO}
\end{equation}
where we choose the harmonic oscillator frequency according to
Eq.~\eqref{eq:omega}, varying the lattice constant $a=\pi/k_c$ and $L=Na$. The
\gls{DVR} method is sometimes referred to as the Lagrange method in numerical
analysis~\cite{Baye:2006}, and functions are usually represented on the spatial
lattice
\begin{align}
  \psi(x) &= \sum_k a \psi(x_k)f_k(x),& 
  \langle f_k|f_l\rangle &= \delta_{kl}.
  \label{eq:psi}
\end{align}
Potential matrix elements usually have a simple and unexpectedly accurate
representation (quasi-locality)
\begin{equation}
  \langle f_k|V|f_l\rangle = \int dx f^*_k(x)V(x)f_l(x) 
  \approx V(x_k)\delta_{kl},
  \label{eq:V}
\end{equation}
where the functions $f_k(x)$ are a linear combination of plane-waves and form an
orthonormal set (these formulae apply for even numbers of abscissa as required
by efficient implementations of the \gls{FFT})
\begin{align}
  f_k(x_l) &= \!\!\!\sum_{n=-N/2}^{N/2-1} 
  \frac{1}{L} \exp \frac{ip_n(x_l-x_k)}{\hbar}\nonumber\\
  & = \begin{cases}
    \frac{\sin\pi(k-l)}{Na}\cot\frac{\pi (k-l)}{N}=0 & k\neq l,\\ 
    1/a & k =l,
  \end{cases} \label{eq:f}\\
  \psi(x_k) &= \sum_l  a f_k(x_l) \psi(x_l),
\end{align}
where $x_k$ and $p_n$ were defined in Eq.~\eqref{eq:pn}.  As before, the
functions $f_k(x_l)$ are simply the normalized projections of the periodic Dirac
functions on the \gls{DVR} subspace \cite{Littlejohn:2002, LCCMP:2002, LC:2002},
and satisfy
\begin{equation}
  \sum_n a f_k(x_n)f_l(x_n) = \delta_{kl}.
\end{equation}  
The sinc-function basis (\ref{eq:sinc}) is obtained in the limit $N\rightarrow
\infty$ (if $a=1$). Similar formulas exist for the calculation of various other spatial derivatives.


While the potential matrix is diagonal, the \gls{DVR} kinetic energy is a matrix
in coordinate representation:
\begin{gather}
  K_{kl} = \begin{cases}
    \frac{\hbar^2\pi^2}{mN^2a^2} 
    \frac{ (-1)^{k-l} }
         { \sin^2 \frac{\pi(k-l)}{N} } & k \neq l\\
    \frac{\hbar^2\pi^2}{6ma^2}\left ( 1+ \frac{2}{N^2}\right ) & 
    k = l.
  \end{cases}
  \label{eq:DVR_K}
\end{gather}
This matrix is full matrix in 1\mysc{d}, but sparse in 3\mysc{d} where only
$1/N^2$ of the matrix elements are non-vanishing. The \gls{HO}
Hamiltonian~\eqref{eq:H_HO} is thus represented in the \gls{DVR} basis with
periodic boundary conditions as
\begin{gather}
  H_{kl} = K_{kl} + \frac{m\omega^2 a^2 k^2}{2}\delta_{kl}.
  \label{eq:dHO}
\end{gather}
The implementation of Dirichlet boundary conditions uses the $\nu=0$ Bessel
function basis (see the \gls{MATLAB} code \cite{matlab} for $l=0$ and also
Ref.~\cite{Baye:2006} for other possible \gls{DVR} basis sets in 1\mysc{d}).

\begin{figure}[tb]
  \includegraphics[width=\columnwidth]{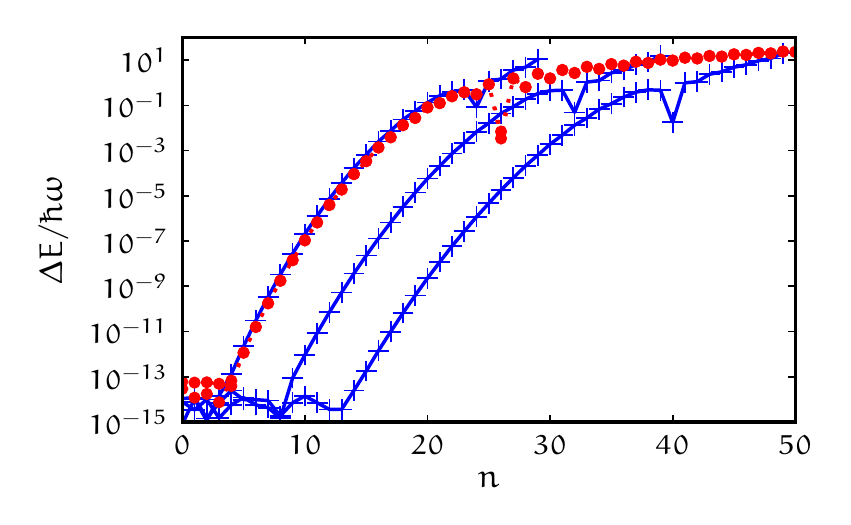}
  \caption{(color online) Difference in spectrum between the \gls{DVR}
    Hamiltonian (\ref{eq:dHO}) and the \gls{HO} energies
    $(n+1/2)\hbar\omega$. The three (blue) curves with pluses have fixed
    \gls{UV} scale (lattice constant $a=1$, $k_c=\pi/a$) with $L=\in\{30, 40,
    50\}$ and $\omega = 2\pi/L$ from left to right.  The (red) curves with dots
    have fixed $L=30$ but varying lattice constant $a \in \{1/2, 1/3\}$
    demonstrating the \gls{UV} convergence. The sizes of the \gls{DVR} basis
    sets are $Lk_c/\pi = 30$, $40$, and $50$ (blue pluses) and $60$, and $90$
    (red circles) respectively. For the blue pluses, the corresponding number of
    harmonic oscillator wave functions suggested in Refs.~\cite{Furnstahl:2012,
      Coon:2012} (see also Eqs.~(\ref{eq:LP})), would be $N = Lk_c /4= L\pi
    /4a\approx $ 24, 31, 39; and 47 and 71 for the red dots, respectively.
    Notice that the size of the \gls{DVR} basis set can be reduced by factor of
    2 to $Lk_c/2\pi = $ 15, 20, 25 (blue) and 30, 45 (red) respectively, by
    imposing Dirichlet boundary conditions, however, in that case, states not
    localized to a single cell will not be reproduced.}
  \label{fig:1}
\end{figure}

\begin{figure}[tb]
  \includegraphics[width=\columnwidth]{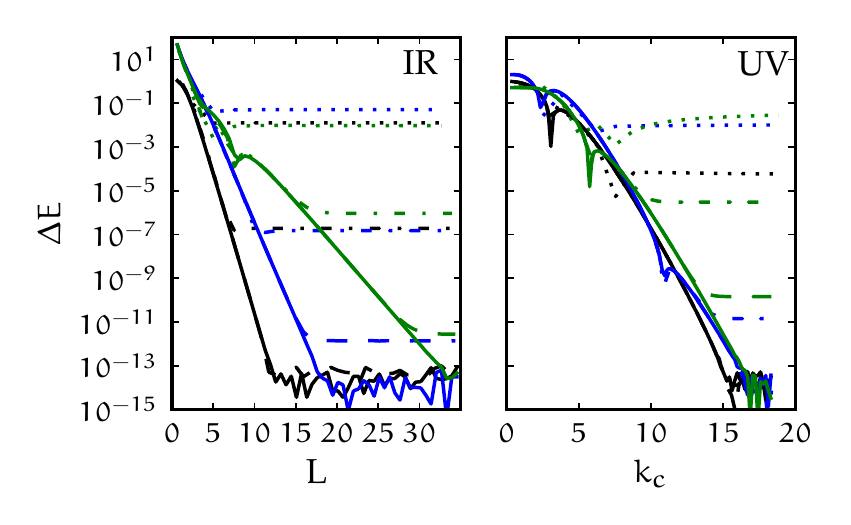}
  \caption{(color online) Exponential convergence of the periodic \gls{DVR}
    basis for the energy of the bound states of the analytically solvable Scarf
    II potential $V(x) = [a+b\sinh x]/\cosh^2x$ (with $\hbar=m=1$).  For $a=7/2$
    and $b=-11/2$, the potential has three bound states -- $E_n = -(3-n)^2/2$
    (shown in black, blue, and green from left to right respectively).  The left
    plot demonstrates the \gls{IR} convergence for increasing $L$ with fixed
    $k_c$; the right plot demonstrates the \gls{UV} convergences for increasing
    $k_c$ for fixed $L$. The various values for $k_c \in \{5, 10, 15, 20\}$
    (left) and $L \in \{5, 15, 25, 35\}$ (right) correspond to dotted,
    dot-dashed, dashed, and solid lines with increasing convergence
    respectively. }
  \label{fig:2}
\end{figure}

In Fig.~\ref{fig:1} we show the energy differences between the eigenvalues of
the Hamiltonian (\ref{eq:dHO}) and $\hbar\omega (n+1/2)$. These ``errors'' are
indicative only of the energy shifts due to the tunneling between neighbouring
cells in the case of periodic boundary conditions, as one can judge by comparing
systems with different lengths at the same energy, when the tunneling matrix
elements are similar. The results for the lowest 2/3 of the spectrum are, for
all practical purposes, converged in the \gls{DVR} method, and the harmonic
oscillator basis set is worse in this case. With $N=Lk_c/4 \approx 24$ one can
obtain at most 10 states or so with a reasonable accuracy in this reduced
interval on the $x$-axis with periodic or Dirichlet boundary conditions, if one
were to follow the prescription of Refs.~\cite{Furnstahl:2012, Coon:2012}.

In Fig.~\ref{fig:2} we demonstrate the \gls{UV} and \gls{IR} exponential
convergence of the \gls{DVR} method for an asymmetric short-range potential with
analytic wavefunctions. Note that both \gls{IR} and \gls{UV} errors scale
exponentially until machine precision is achieved -- $\Delta E \propto
\exp\bigl(-2k(L) L\bigr)$ (\gls{IR}) and $\Delta E \propto \exp(-2k_c r_0)$
(\gls{UV}) respectively, where $r_0$ is potential dependent and $k(L)$ is
determined by the bound state energy $E(L) = -\hbar^2k^2(L)/2m$.  These
exponential scalings follow from simple Fourier analysis (\gls{UV}) and band
structure theory (\gls{IR}) for short-ranged smooth potentials.  Note in
particular that the linear \gls{UV} scaling differs from the quadratic empirical
dependence discussed in~\cite{Furnstahl:2012}.  We have also demonstrated the
utility of the \gls{DVR} method for a variety of \gls{DFT} and \gls{QMC}
many-body calculations.

The Bessel-function \gls{DVR} basis $j_l(\Lambda r_n/\hbar)$ for spherical
coordinates was used in~\cite{Bulgac:2007a, matlab} to solve the
self-consistent \gls{SLDA} \gls{DFT} equations for the harmonically trapped
unitary Fermi gas.  While the basis is defined for all $l$, even and odd
$l$-partial radial wave functions can be effectively expressed using only the
$j_0$ and $j_1$ basis sets respectively (see~\cite{Nygaard:2004}) with the
angular coordinates represented by spherical harmonics.  The spatial mesh size
is given by $\Delta r = r_{n+1}-r_{n} \approx \pi \hbar / \Lambda$.  Applied to
nuclear matter, $\Lambda= \SI{600}{MeV}/c$ gives $\Delta r \approx \SI{1}{fm}$
and $N_s = R/\Delta r \approx 20$ radial mesh points in a spherical box of
radius $R\approx\SI{20}{fm}$. A \gls{MATLAB} code for a spin imbalanced trapped
unitary gas with pairing and using two different chemical potentials for the
spin-up and spin-down fermions respectively, is about \num{400} lines and
converges in a few seconds on a laptop~\cite{matlab, plots}.

The periodic \gls{DVR} basis was used in Ref.~\cite{BF:2008} to solve the
self-consistent \gls{SLDA} \gls{DFT}, predicting a supersolid \gls{LO} phase in
the spin imbalanced unitary Fermi gas.  Explicit summation over Bloch momenta
was used to remove any \gls{IR} errors (i.e. simulating a periodic state in
infinite space rather than in a periodic space.)  The periodic basis was also
used in~\cite{Bulgac:2009} to demonstrate the Higgs mode by solving the
time-dependent \gls{SLDA} for systems with up to $10^5$ particles. (In both
these approaches, spatial variations were only allowed in one direction:
transverse directions were treated analytically.)

Full 3\mysc{d} periodic \gls{DVR} bases were used in~\cite{Bulgac:2011c} to
solve the time-dependent \gls{SLDA} equations for $48\times 48 \times 48$ and
$196 \times 32 \times 32$ lattices, solving $\approx 5\times 10^5$ non-linearly
coupled partial-differential equations for several million time steps to study
the real-time dynamics of the superfluid unitary Fermi gas.  Extensions of this
code on current supercomputers allow us to increase the overall size of such
problems by an order of magnitude.  These 3\mysc{d} \gls{DVR} bases were also
used to study the \gls{GDR} in deformed triaxial open-shell heavy
nuclei~\cite{Stetcu:2011} without any symmetry restrictions.  Finally, the
3\mysc{d} \gls{DVR} basis was used in~\cite{Wlazlowski:2013, *Drut:2012a} (and
earlier references therein) to perform \textit{ab initio} \gls{QMC} calculations
of strongly interacting fermions in spatial lattices ranging from
$6^3=\num{216}$ to $16^3=\num{4096}$ for systems comprising \num{20} to
\num{160} particles and with \num{5000} steps in imaginary time.  These systems
are significantly larger than the $364$ single-particle states used
in~\cite{Gilbreth:2013} to implement a nuclear shell-model
\gls{QMC}~\cite{Koonin:1997}.  Similar applications of \gls{DVR} \gls{QMC} are
currently being developed for nuclear systems.

\begin{figure}
  \includegraphics[width=\columnwidth]{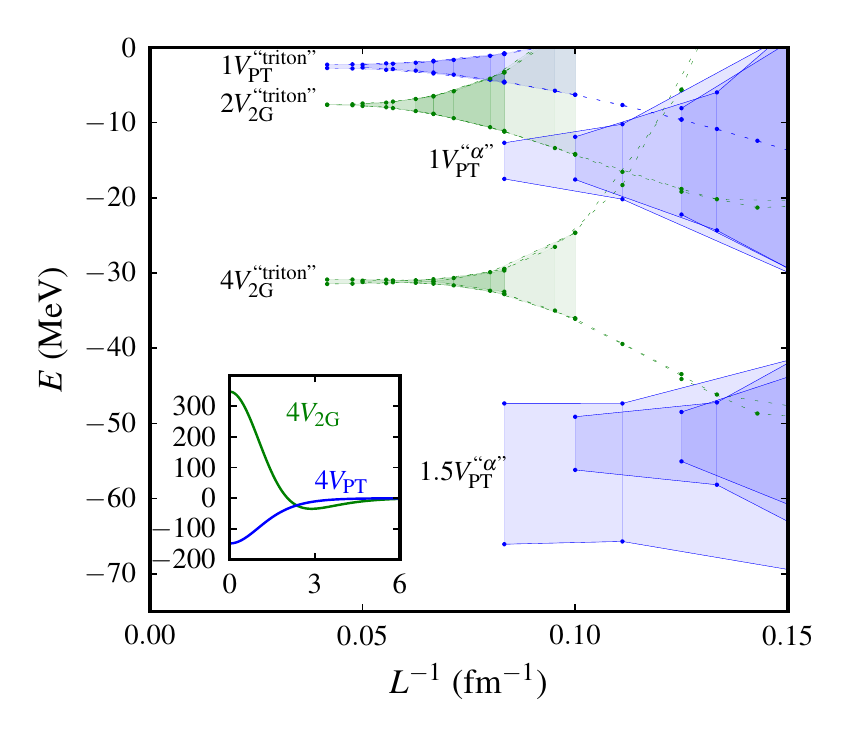}
  \caption{(color online) Binding energy of a three-particle ``triton'' and
    four-particle ``alpha`` ground state using various multiples (specified by
    numerical factors in the figure) of the potentials $V_{\text{PT}}(r) \propto
    -\smash{\sech^2}(2r/r_0)$ and $V_{2\text{G}}(r) \propto \exp(-r^2/r_0^2) -
    4\exp(-4r^2/r_0^2)$ with $r_0=\SI{3}{fm}$ (see~\cite{Forbes:2012} for
    explicit normalizations).  Upper and lower bounds are obtained from
    Dirichlet and periodic boundary conditions respectively.  The deeply bound
    four-body state with $1.5V_{\text{PT}}$ is not converged and has comparable
    \gls{UV} and \gls{IR} errors (each ``band'' has fixed lattice spacing).  The
    other results are \gls{UV} converged: the different lattice spacings lying
    on the same curves describing the dependence on the box size.  The inset
    shows the radial profile of the two potentials.}
  \label{fig:3}
\end{figure}

We further illustrate the power of the \gls{DVR} basis in Fig.~\ref{fig:3} by
solving the 6\mysc{d} and 9\mysc{d} Schrödinger equations for three-body
(``triton'') and four-body (``$\alpha$'') bound states with distinguishable
particles interacting with two centrally symmetric potentials: a purely
attractive Posh-Teller potential, and an attractive potential with a repulsive
core (see the inset).  We used a Cartesian lattice for the relative Jacobi
coordinates to eliminate the center-of-mass coordinate.  Our goal was to solve
these with a modern laptop (\SI{2.7}{G\hertz} Intel Core i7 MacBook Pro with
\SI{16}{GB} of \gls{RAM}) in no more than about a few minutes, without any
tricky optimizations such as taking advantage of symmetry properties of the
wavefunction.  (Parity alone could reduce the Hilbert space by factors of $2^6$
and $2^9$ respectively.)  Coding these problems is simple -- the \gls{MATLAB}
versions are about \num{200} lines per problem while the general Python code is
about \num{1000} lines (including documentation and tests) \cite{plots}.  We are
not aware of other attempts to solve directly the Schrödinger equation in a
9\mysc{d}-space.

To compute the ground state energy, we use two alternative techniques: imaginary
time evolution of a trial state (slow convergence but gives a representative
wavefunction) and a simple Lanczos algorithm (fast convergence, but only a few
low-energy eigenvalues).  For the triton we used lattices $N_s^6=8^6\cdots
16^6$: for the $\alpha$ state we use lattices $N_s^9=4^9\cdots 8^9$. The size of
the largest Hilbert space is thus $\approx 1.68\times 10^7$ for the triton and
$8^9\approx 1.34\times 10^8$ for the $\alpha$.  Several spatial mesh sizes $a =
$\SIrange{0.5}{1.5}{fm} corresponding to $\Lambda \approx$
\SIrange{300}{930}{MeV}$/c$ are used to explore convergence.  Note that, unlike
with other methods used for nuclear structure calculations, adding local
three-body and four-body interaction will neither complicate the code nor
impact the performance.

As discussed earlier, the \gls{UV} convergence is determined by the properties
of the interaction: For example, the high-momentum components of a wavefunction
in a short-range potential will have a power-law decay $\propto k^{-4}$
\cite{Sartor:1980, *Tan:2008uq} (rather than an exponential decay). The \gls{IR}
convergence of the energy will be determined by the energy of the lowest
many-body threshold.  For example, if there is an S-wave two-body threshold with
binding energy difference $Q(L)$ in the box, then the \gls{IR} error will
be~\cite{Luscher:1986, *Beane:2003da}
\begin{equation}
  E(L)\approx E_\infty + \frac{A\exp(-\sqrt{2M Q(L)}L/\hbar)}{L}
\end{equation}
where $M$ is the corresponding reduced mass, and $A$ an asymptotic normalization
factor that is positive or negative for Dirichlet or periodic boundary
conditions respectively. If the lowest threshold is higher-body or in a
different (not S-wave) configuration, then this behaviour will be modified in a
straightforward manner.  (Competition between several closely lying thresholds
will further complicate the \gls{IR} convergence properties.) Note that this
differs from the results of~\cite{Furnstahl:2012, Coon:2012}.

In summary, the \gls{DVR} basis seems ideal for nuclear structure calculations
using either \gls{DFT}, \gls{QMC} or configuration mixing approaches.  It is
near optimal in size, and can deliver results with exponential convergence. The
\gls{DVR} basis shares the important advantages of the \gls{HO} basis set:
efficiently separating out the center-of-mass motion using Jacobi coordinates
(with the added benefit of not needing to evaluate Talmi-Moshinsky
coefficients), utilizing symmetries to reduce the basis size (spherical with the
Bessel function \gls{DVR}). Moreover, matrix elements are easy to evaluate --
the potential matrix is diagonal for local potentials (no overlap integrals are
needed -- see for example Eq.~\eqref{eq:V}), the kinetic energy matrix is sparse
and explicitly expressed analytically, and many-body forces can be easily
included.  Furthermore, the \gls{UV} and \gls{IR} convergence properties of the
basis appear on a equal footing, and are clearly expressed in terms of the
momentum-space projection and finite box size, allowing for simplified and sound
convergence analysis, with a clear mathematical underpinning. Finally, we
demonstrated that the \gls{DVR} basis can be used in extremely large Hilbert
spaces with relatively modest computational resources.

\providecommand{\MMFGRANT}{\mysc{de-fg02-00er41132}}

We thank G.F. Bertsch for discussions and the
support under \mysc{us} \gls{DoE} grants \mysc{de-fg02-97er41014},
\mysc{de-fc02-07er41457}, and \MMFGRANT.

%

\end{document}